\documentclass[sigconf]{acmart}
\author{Subinay Adhikary}
 \email{sa21rs094@iiserkol.ac.in}
 \affiliation{%
   \institution{Indian Institute of Science Education and Research Kolkata}
  \city{Kolkata}
   \country{India}
}
\author{Sagnik Das}
 \email{sd21ms065@iiserkol.ac.in}
 \affiliation{%
   \institution{Indian Institute of Science Education and Research Kolkata}
  \city{Kolkata}
   \country{India}
}
\author{Sagnik Saha}
 \email{ss21ms213@iiserkol.ac.in}
 \affiliation{%
   \institution{Indian Institute of Science Education and Research Kolkata}
  \city{Kolkata}
   \country{India}
}

 \author{Procheta Sen}
 \email{procheta.sen@liverpool.ac.uk}
 \affiliation{%
   \institution{University of Liverpool}
   \streetaddress{P.O. Box 1212}
  \city{Liverpool}
   \country{United kingdom}
}
\author{Dwaipayan Roy}
 \email{dwaipayan.roy@iiserkol.ac.in}
 \affiliation{%
   \institution{Indian Institute of Science Education and Research Kolkata}
  \city{Kolkata}
   \country{India}
}
\author{Kripabandhu Ghosh}
 \email{kripaghosh@iiserkol.ac.in}
 \affiliation{%
   \institution{Indian Institute of Science Education and Research Kolkata}
  \city{Kolkata}
   \country{India}
}
\AtBeginDocument{%
  \providecommand\BibTeX{{%
    \normalfont B\kern-0.5em{\scshape i\kern-0.25em b}\kern-0.8em\TeX}}}

\setcopyright{acmcopyright}
\copyrightyear{2018}
\acmYear{2018}
\acmDOI{XXXXXXX.XXXXXXX}

\acmConference[Conference acronym 'XX]{Make sure to enter the correct
  conference title from your rights confirmation emai}{June 03--05,
  2018}{Woodstock, NY}
\acmPrice{15.00}
\acmISBN{978-1-4503-XXXX-X/18/06}

\begin{document}

\title{\textit{A}utomated \textit{A}ttribute \textit{E}xtraction from \textit{Le}gal \textit{P}roceedings}

\renewcommand{\shortauthors}{}

\begin{abstract}
 The escalating number of pending cases is a growing concern worldwide. Recent advancements in digitization have opened up possibilities for leveraging artificial intelligence (AI) tools in the processing of legal documents. Adopting a structured representation for legal documents, as opposed to a mere bag-of-words flat text representation, can significantly enhance processing capabilities. With the aim of achieving this objective, we put forward a set of diverse attributes for criminal case proceedings. We use a state-of-the-art sequence labeling framework to automatically extract attributes from the legal documents. Moreover, we demonstrate the efficacy of the extracted attributes in a downstream task, namely legal judgment prediction.
\end{abstract}


%\ccsdesc[500]{Computer systems organization~Embedded systems}




\maketitle

\section{Introduction}

The exponential increase in the number of pending cases (such as over $10$ million\footnote{\url{https://njdg.ecourts.gov.in/njdgnew/index.php}} pending cases in India) highlights the pressing need for efficient processing of legal documents using AI tools. %In light of this global scenario, leveraging AI technology has become more crucial than ever to streamline and expedite legal processes.
 %The objective of the AI tools is to assist lawyers and common people to effectively execute the law and order process in any country.
%In this work, we specifically focus on the efficient processing of legal documents from a lawyer's perspective. 
%
Legal experts require time to thoroughly review lengthy documents. On average, a case document contains $3,829$ words. Existing research \cite{ANAND20222141} proposed approaches to provide different kinds of summaries for a legal document.

However, a structured representation of a document is better \cite{MUNIR2018116,inbook} in terms of efficiency in processing any downstream task. To address this, we propose to represent a legal document in terms of different important concepts or attributes. By collaborating with legal experts, we formulated a comprehensive set of concepts specifically for criminal case documents. %The motivation of this work is that extracting important concepts from a legal document will improve the performance of different downstream applications in the Law-AI domain.
Law students were appointed to annotate $200$ legal documents (significant annotations compared to other legal works in India \citeauthor{BhattacharyaPG019,malik2021ildc})  
with $7$ different attributes.
%We used law students to annotate a set of documents in the gold standard. 
However, it is not possible to manually annotate a large number of documents. Consequently, we explore existing sequence labeling approaches using deep neural networks to effectively extract the concepts or attributes from legal documents. 

%Broadly speaking, sequence labeling approaches are used for entity or attribute extraction for various domains. 
%
Broadly speaking, the contributions of this paper are as follows.
\begin{enumerate}
    
\item  We propose a new framework to present each legal document in terms of different important concepts or attributes.\\
\item  We show the effectiveness of existing sequence labeling approaches in automatically extracting attributes from legal documents using a limited amount of training data.\\
\item We also demonstrate the effectiveness of structured representation in a downstream task (i.e. judgment prediction).

\end{enumerate}

\section{Related Work}
Existing literature related to our research scope can be broadly categorized into three areas: a) Legal NLP b) Entity Recognition. Each one of them is described as follows. 

\textbf{Legal AI}
In legal cases, the documents often encompass lengthy and intricate sentences, making it challenging and time-consuming to thoroughly read and comprehend the entire content of a case document. To alleviate this extensive effort, researchers have emphasized the extraction of noun phrases known as \textit{catchphrases }\cite{catchphrase1, catchphrase2} from the document. This approach aims to capture the key elements and central themes of the text. Additionally, \textit{summarization} \cite{ANAND20222141,shukla-etal-2022-legal} techniques have been proven to be effective in gaining a comprehensive understanding of the document by condensing its content into a concise summary. However, to get the thematic view of the document, we need to extract fine-grained level information or concepts from the document. We plan to capture the thematic view of a legal document in our proposed attribute framework for legal documents.
\begin{figure*}[htb]
\center
    \includegraphics[width=0.7\textwidth]{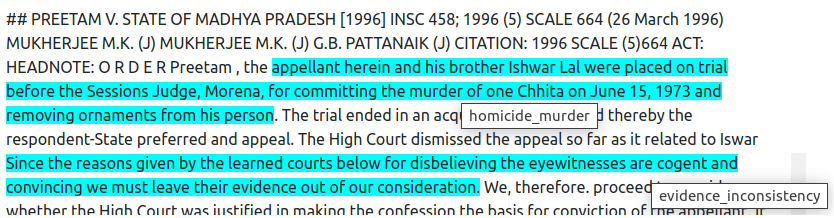}
    \caption{Annotation of a Legal Document}
    \label{fig:anno}
\end{figure*}

\textbf{Entity Recognition}
Named Entity recognition \cite{entity} is a well-established task in NLP. Entities are defined as objects which have an independent existence. Examples of named entities are persons and organizations. Domain-specific entities (e.g. biomedical entity) are in general defined by domain experts through an ontology structure. Initially rule-based systems \cite{rule_entity}, and syntactical structure \cite{syntactic} were used to automatically extract entities from a text. With the advancement of machine learning and deep learning techniques existing research has used conditional random fields, sequence-to-sequence models, and large language models for \textit{Named entity recognition} \cite{entity1,entity2,entity3,entity4}. More specifically \textit{legal name entities} extraction \cite{skylaki2020named,kalamkar2022named} has been implemented using Conditional Random Fields (CRF %\citeauthor{lafferty2001conditional}) 
by %\citeauthor{cardellino2017low} (2017). \citeauthor{tran2018automatic} (2018) and \citeauthor{truong2017single} (2017) 
used a sequence tagger, specifically the BiLSTM-CRF model. To the best of our knowledge, this is the initial endeavor to automate attribute extraction from legal proceedings. \\ 
%\textbf{Novelity of this work:}
%Catch Phrase Extraction 
%Summarization 

\section{Automatic Attribute Extraction}\label{sec:annotate}
%Here we first describe the process of annotation of legal documents with different attributes. Consequently, we describe the automatic annotation process in the context of legal documents.
%\subsection{Legal Document Annotation Framework}
In this work, we particularly focused on criminal case documents from the Supreme Court of India\footnote{Available from \url{https://indiankanoon.org/}}. We propose seven important legal concepts or attributes corresponding to a criminal case document with the help of law experts. The different concepts and the corresponding descriptions are as follows.
\begin{enumerate}
    \item  \textbf{ExpertWittest:} Witness testimony from Expert people includes forensic and ballistic experts. 
    \item \textbf{Wittest:} Testimony from non-Expert witness testimony has been mentioned during the judgment.
    \item \textbf{Assault:} Hurt by a sharp weapon.
    \item \textbf{Riot:} Unlawful enterprise in a violent manner.
    \item \textbf{Homicide:} Homicide amounting to murder.
    \item \textbf{Imprisonment:} Sentenced to life imprisonment.
    \item \textbf{Evidence:} Evidence of the crime was  found.
\end{enumerate}

 In each document, legal experts highlighted the text span corresponding to a concept present in that legal document. Figure \ref{fig:anno} shows an example annotation. Legal experts annotated $200$ legal documents with gold standard annotations. 
 % Figure \ref{fig:anno} shows an example gold standard annotation.

\subsection{Problem Definition}
It is not possible to manually annotate all the legal documents. However, to exploit the attributes corresponding to each case in a downstream task we would need a lot of annotated documents. As a result of this, we focus on the automatic annotation of legal documents with the attributes or concepts.
We cast the automatic annotation problem as a sequence labeling problem \cite{rei-sogaard-2018-zero}. In this context, each token in the input text is assigned a corresponding label. Sequence labeling has been extensively employed in NLP research to tackle various challenges.
The advantage of casting the automatic annotation of a document into a sequence labeling problem is that along with sequence labeling along with all the attributes mentioned in a case, we can also obtain the highlighted sequence of tokens responsible for generating that attribute.

As described in Section \ref{sec:annotate}, there are 
$7$ important concepts or labels in a legal document. All the tokens in a legal document may not be useful for annotation. Consequently, we introduced an additional attribute named `NoTag' to complement the aforementioned seven concepts (i.e., $8$ in total).

\subsection{Methodology}
Here we describe the working principle of a sequence labeling model. If $x$ is the input sequence (i.e. $x=x_1,x_2,\ldots x_n$) and $y$ is the desired output sequence (i.e. $y=y_1,y_2,\ldots y_{n'}$) then $n=n'$. The sequence labeling problem learns a mapping $H : X \xrightarrow{}  Y$ using a state-of-the-art BiLSTM-CRF \cite{huang2015bidirectional}. The BiLSTM-CRF network essentially computes the maximum likelihood of each sequence label as follows.
\begin{equation}
    P(y_{0,\ldots n}|r_{0 \ldots n})=\prod_{i=1}^n f(y_{i-1}, y_{i}, x)
    \label{bilstm-crf}
\end{equation}
In Equation \ref{bilstm-crf}, $f(y_{i-1}, y_{i}, x)$ estimates the likelihood of an individual label (i.e. $y_i$) based on its neighboring sequence. The hidden state output vectors ($r_i$) obtained from a BiLSTM model are used to estimate the likelihood in $f(y_{i-1}, y_{i}, x)$.

\section{Experiment Framework}
\textbf{Dataset:} For sequence labeling input only the highlighted spans from the annotators are used for training. Two annotators manually annotated $200$ documents. While highlighting a span, annotators were only asked to choose the sequence of words which are relevant. The highlighted text may not span over a whole sentence or it may span across a sentence. The details of the dataset are shown in Table \ref{tab:dataset}.

For judgment prediction downstream task, we used a dataset\footnote{\url{https://github.com/Exploration-Lab/CJPE}} consisting of $5,082$ documents for training and $94$ documents for testing.
\begin{table}[t]
    \footnotesize
    \resizebox{\columnwidth}{!}{
    \begin{tabular}{|l|c|c|c|c|c|c|c|c|}
    \hline
    \multicolumn{2}{|c|}{Dataset} &\multicolumn{7}{|c|}{\textbf{ Tags} }\\
 % & \multicolumn{8}{|c|}{\textbf{ Accuracy}} \\
    \hline
  Type& Statistics & ExpWittet & Wittest & Homicide & Assault & Imprisonment & Riot & Evidence \\  
 %   \textbf{Method} & \textbf{ExpWitTest} & \textbf{WitTest} &  \textbf{Homicide} & \textbf{Assault} & \textbf{Imprisonment}& \textbf{Riot}& \textbf{Evidence}& \textbf{Overall}\\
    \hline
  Train & \#sentences & 31 & 35 & 50 & 39 & 50 & 32 & 24  \\
     \hline
    Train & \#tokens & 502 & 930 & 501 & 734 & 509 & 797 & 580 \\
\hline
    \hline
     Test  &\#sentences & 372 & 162 & 78 & 48 & 34 & 20 & 10 \\
       \hline
     Test  &\#tokens & 7789 & 3841 & 1036 & 755 & 344 & 410 & 257 \\
 \hline
    \end{tabular}}
 \caption{Dataset Statistics}
\label{tab:dataset}
 \end{table}

\begin{table}[t]
    \footnotesize
    \resizebox{\columnwidth}{!}{
    \begin{tabular}{|l|c|c|c|c|c|c|c|c|}
    \hline
  & \multicolumn{8}{|c|}{\textbf{ Accuracy}} \\
    \hline
    
    \textbf{Method} & \textbf{ExpWitTest} & \textbf{WitTest} &  \textbf{Homicide} & \textbf{Assault} & \textbf{Imprisonment}& \textbf{Riot}& \textbf{Evidence}& \textbf{Overall}\\
    \hline
   SeqLabel (Roberta) & \textbf{0.85} & 0.50 & 0.18 & \textbf{0.39} & 0.40& \textbf{0.78} & \textbf{0.24} & 0.67 \\
 SeqLabel (BERT) & 0.61 & 0.28 & 0.14 & 0.35 & 0.36 & 0.41 & 0.18 & 0.60 \\
 SeqLabel (LegalBERT) & 0.84 & \textbf{0.51} &\textbf{0.35}  & 0.36 &\textbf{0.50}  & 0.69 & 0.15 & \textbf{0.68} \\
 \hline
    \end{tabular}}
    \caption{Automated Tag Extraction Results using Flair}
\label{tab:ubar}
 \end{table}

\textbf{Preprocessing}
As a preprocessing step, each legal case document was split into sentences using a sentence tokenizer from NLTK \footnote{\url{https://www.nltk.org/api/nltk.tokenize.html}}. Then each highlighted sentence was split into tokens based on whitespace. While preparing the dataset for the training of the sequence labeling model, the words not highlighted within a sentence are tagged with  \texttt{NoTag}.
The sequence labeling approach is implemented in Flair model\footnote{\url{https://pypi.org/project/flair/}} and is used for all our experiments.

 %\subsubsection{Judgement Prediction:}
As suggested in \cite{, malik2021ildc} we utilized BERT-uncased to generate embeddings for the last $510$ tokens of each document for judgment prediction task. The embeddings were eventually used as input to a logistic regression classifier. %The accuracy achieved in this phase was $0.53$.
%In the second part, 
For the `with tag' version in Table \ref{tab:judgement}, we appended all the extracted attributes of each test document to its corresponding content. Table \ref{tab:judgement} shows the method with tag overperforms the without tag version.
\begin{table}[htb]
    \footnotesize
    \begin{tabular}{|l|c|c|c|c|c|c|c|}
    \hline
& & \multicolumn{5}{|c|} {Metric} \\
    \hline
 \multicolumn{2}{|c|} {Method}    & \multicolumn{2}{|c|} {Acc (Class 0)} & \multicolumn{2}{|c|} {Acc (Class 1)} &\\
    \hline
   Embedding  & Input Format & Precision& Recall & Precision & Recall& Acc \\
    \hline
    BERT  & Text & 0.50 & 0.34 &  0.54 & 0.70 & 0.53 \\
    \hline
     BERT& Text+Tag &  0.54 & \textbf{0.52} & 0.59  & 0.62 & 0.56 \\
    \hline
      BERT &Text+Span & 0.61 & 0.25 & 0.56 & 0.86 &  0.57\\
    \hline
   InLegalBERT &Text & 0.57 & 0.36 & 0.57 & 0.76 & 0.57 \\
    \hline
     InLegalBERT & Text + Tag & 0.63 & 0.31 & 0.58 & 0.84 & 0.60 \\   
       \hline
      InLegalBERT &  Text +Span & \textbf{0.75} & 0.40 & \textbf{0.62} & \textbf{0.88} & \textbf{0.66} \\
    \hline
\end{tabular}
\caption{Judgement Prediction Results}
\label{tab:judgement}
\end{table}

\section{Results}
We can observe from Table \ref{tab:ubar} that from all the different variations of the Flair model, the Roberta version (i.e. SeqLabel (LegalBERT)) performed the best in terms of overall accuracy. The accuracy for \texttt{NoTag} was more than $80\%$ for all the approaches. Consequently, we didn't report it in Table \ref{tab:ubar}.

To investigate the effectiveness of the legal document attributes in a downstream task, we explored both `Text+Tag' and `Text+span' approaches in Table \ref{tab:judgement}. For the `Text+Tag' version, we appended each test document's extracted attributes to its corresponding content. Similarly for the `Text+Span' version, we appended the highlighted texts corresponding to each tag to the content of the document. Table \ref{tab:judgement} shows that for both BERT and InLegalBERT variations the `Text+Tag' and `Text+Span' versions performed better compared to using only the text version. It is also interesting to observe that the `Text+Span' version is performing better compared to the `Text+Tag' version. Consequently, we can conclude that the highlighted portions corresponding to each tag are more meaningful in legal document representation compared to the tag names.
\section{Conclusion}

In this paper, we propose a novel attribute framework to capture the thematic view of a legal document in a structured way. We also show the effectiveness of a state-of-the-art sequence labeling approach in automatically predicting the attributes of a legal document. We also demonstrate the usefulness of the attributes in a downstream task. It is always difficult to manually annotate a reasonable number of training data for a new task. In the future, we would like to explore advanced deep-learning approaches to address the issue of not having enough training data for automatic attribute extraction model training.
\bibliographystyle{acm-reference-format}
\bibliography{ref}


\begin{thebibliography}{19}

%%% the Web address.

\ifx \showCODEN    \undefined \def \showCODEN     #1{\unskip}     \fi
\ifx \showDOI      \undefined \def \showDOI       #1{#1}\fi
\ifx \showISBNx    \undefined \def \showISBNx     #1{\unskip}     \fi
\ifx \showISBNxiii \undefined \def \showISBNxiii  #1{\unskip}     \fi
\ifx \showISSN     \undefined \def \showISSN      #1{\unskip}     \fi
\ifx \showLCCN     \undefined \def \showLCCN      #1{\unskip}     \fi
\ifx \shownote     \undefined \def \shownote      #1{#1}          \fi
\ifx \showarticletitle \undefined \def \showarticletitle #1{#1}   \fi
\ifx \showURL      \undefined \def \showURL       {\relax}        \fi
% invisible to TeX
\providecommand\bibfield[2]{#2}
\providecommand\bibinfo[2]{#2}
\providecommand\natexlab[1]{#1}
\providecommand\showeprint[2][]{arXiv:#2}

\bibitem[Anand and Wagh(2022)]%
        {ANAND20222141}
\bibfield{author}{\bibinfo{person}{Deepa Anand} {and} \bibinfo{person}{Rupali
  Wagh}.} \bibinfo{year}{2022}\natexlab{}.
\newblock \showarticletitle{Effective deep learning approaches for
  summarization of legal texts}.
\newblock \bibinfo{journal}{\emph{Journal of King Saud University - Computer
  and Information Sciences}} \bibinfo{volume}{34}, \bibinfo{number}{5}
  (\bibinfo{year}{2022}), \bibinfo{pages}{2141--2150}.
\newblock
\showISSN{1319-1578}
\urldef\tempurl%
\url{https://doi.org/10.1016/j.jksuci.2019.11.015}
\showDOI{\tempurl}


\bibitem[Bhattacharya et~al\mbox{.}(2019)]%
        {BhattacharyaPG019}
\bibfield{author}{\bibinfo{person}{Paheli Bhattacharya},
  \bibinfo{person}{Shounak Paul}, \bibinfo{person}{Kripabandhu Ghosh},
  \bibinfo{person}{Saptarshi Ghosh}, {and} \bibinfo{person}{Adam Wyner}.}
  \bibinfo{year}{2019}\natexlab{}.
\newblock \showarticletitle{Identification of Rhetorical Roles of Sentences in
  Indian Legal Judgments}. In \bibinfo{booktitle}{\emph{Legal Knowledge and
  Information Systems - {JURIX} 2019: The Thirty-second Annual Conference,
  Madrid, Spain, December 11-13, 2019}} \emph{(\bibinfo{series}{Frontiers in
  Artificial Intelligence and Applications}, Vol.~\bibinfo{volume}{322})},
  \bibfield{editor}{\bibinfo{person}{Michal Araszkiewicz} {and}
  \bibinfo{person}{V{\'{\i}}ctor Rodr{\'{\i}}guez{-}Doncel}} (Eds.).
  \bibinfo{publisher}{{IOS} Press}, \bibinfo{pages}{3--12}.
\newblock
\urldef\tempurl%
\url{https://doi.org/10.3233/FAIA190301}
\showDOI{\tempurl}


\bibitem[Curran and Clark(2003)]%
        {entity1}
\bibfield{author}{\bibinfo{person}{James~R. Curran} {and}
  \bibinfo{person}{Stephen Clark}.} \bibinfo{year}{2003}\natexlab{}.
\newblock \showarticletitle{Language Independent NER using a Maximum Entropy
  Tagger}. In \bibinfo{booktitle}{\emph{Proceedings of CoNLL-2003}},
  \bibfield{editor}{\bibinfo{person}{Walter Daelemans} {and}
  \bibinfo{person}{Miles Osborne}} (Eds.). \bibinfo{publisher}{Edmonton,
  Canada}, \bibinfo{pages}{164--167}.
\newblock


\bibitem[Hammerton(2003)]%
        {entity2}
\bibfield{author}{\bibinfo{person}{James Hammerton}.}
  \bibinfo{year}{2003}\natexlab{}.
\newblock \showarticletitle{Named Entity Recognition with Long Short-Term
  Memory}. In \bibinfo{booktitle}{\emph{Proceedings of CoNLL-2003}},
  \bibfield{editor}{\bibinfo{person}{Walter Daelemans} {and}
  \bibinfo{person}{Miles Osborne}} (Eds.). \bibinfo{publisher}{Edmonton,
  Canada}, \bibinfo{pages}{172--175}.
\newblock


\bibitem[Horrocks(2013)]%
        {inbook}
\bibfield{author}{\bibinfo{person}{Ian Horrocks}.}
  \bibinfo{year}{2013}\natexlab{}.
\newblock \bibinfo{booktitle}{\emph{What Are Ontologies Good For?}}
\newblock \bibinfo{pages}{175--188}.
\newblock
\showISBNx{978-3-642-34996-6}
\urldef\tempurl%
\url{https://doi.org/10.1007/978-3-642-34997-3_9}
\showDOI{\tempurl}


\bibitem[Huang et~al\mbox{.}(2015)]%
        {huang2015bidirectional}
\bibfield{author}{\bibinfo{person}{Zhiheng Huang}, \bibinfo{person}{Wei Xu},
  {and} \bibinfo{person}{Kai Yu}.} \bibinfo{year}{2015}\natexlab{}.
\newblock \bibinfo{title}{Bidirectional LSTM-CRF Models for Sequence Tagging}.
\newblock
\newblock
\showeprint[arxiv]{1508.01991}~[cs.CL]


\bibitem[Kalamkar et~al\mbox{.}(2022)]%
        {kalamkar2022named}
\bibfield{author}{\bibinfo{person}{Prathamesh Kalamkar}, \bibinfo{person}{Astha
  Agarwal}, \bibinfo{person}{Aman Tiwari}, \bibinfo{person}{Smita Gupta},
  \bibinfo{person}{Saurabh Karn}, {and} \bibinfo{person}{Vivek Raghavan}.}
  \bibinfo{year}{2022}\natexlab{}.
\newblock \showarticletitle{Named entity recognition in indian court
  judgments}.
\newblock \bibinfo{journal}{\emph{arXiv preprint arXiv:2211.03442}}
  (\bibinfo{year}{2022}).
\newblock


\bibitem[Malik et~al\mbox{.}(2021)]%
        {malik2021ildc}
\bibfield{author}{\bibinfo{person}{Vijit Malik}, \bibinfo{person}{Rishabh
  Sanjay}, \bibinfo{person}{Shubham~Kumar Nigam}, \bibinfo{person}{Kripa
  Ghosh}, \bibinfo{person}{Shouvik~Kumar Guha}, \bibinfo{person}{Arnab
  Bhattacharya}, {and} \bibinfo{person}{Ashutosh Modi}.}
  \bibinfo{year}{2021}\natexlab{}.
\newblock \showarticletitle{ILDC for CJPE: Indian legal documents corpus for
  court judgment prediction and explanation}.
\newblock \bibinfo{journal}{\emph{arXiv preprint arXiv:2105.13562}}
  (\bibinfo{year}{2021}).
\newblock


\bibitem[Mandal et~al\mbox{.}(2017)]%
        {catchphrase2}
\bibfield{author}{\bibinfo{person}{Arpan Mandal}, \bibinfo{person}{Kripabandhu
  Ghosh}, \bibinfo{person}{Arindam Pal}, {and} \bibinfo{person}{Saptarshi
  Ghosh}.} \bibinfo{year}{2017}\natexlab{}.
\newblock \showarticletitle{Automatic Catchphrase Identification from Legal
  Court Case Documents}. In \bibinfo{booktitle}{\emph{Proceedings of the 2017
  ACM on Conference on Information and Knowledge Management}} (Singapore,
  Singapore) \emph{(\bibinfo{series}{CIKM '17})}.
  \bibinfo{publisher}{Association for Computing Machinery},
  \bibinfo{address}{New York, NY, USA}, \bibinfo{pages}{2187–2190}.
\newblock
\showISBNx{9781450349185}


\bibitem[Munir and {Sheraz Anjum}(2018)]%
        {MUNIR2018116}
\bibfield{author}{\bibinfo{person}{Kamran Munir} {and} \bibinfo{person}{M.
  {Sheraz Anjum}}.} \bibinfo{year}{2018}\natexlab{}.
\newblock \showarticletitle{The use of ontologies for effective knowledge
  modelling and information retrieval}.
\newblock \bibinfo{journal}{\emph{Applied Computing and Informatics}}
  \bibinfo{volume}{14}, \bibinfo{number}{2} (\bibinfo{year}{2018}),
  \bibinfo{pages}{116--126}.
\newblock
\showISSN{2210-8327}


\bibitem[Rei and S{\o}gaard(2018)]%
        {rei-sogaard-2018-zero}
\bibfield{author}{\bibinfo{person}{Marek Rei} {and} \bibinfo{person}{Anders
  S{\o}gaard}.} \bibinfo{year}{2018}\natexlab{}.
\newblock \showarticletitle{Zero-Shot Sequence Labeling: Transferring Knowledge
  from Sentences to Tokens}. In \bibinfo{booktitle}{\emph{Proceedings of the
  2018 Conference of the North {A}merican Chapter of the Association for
  Computational Linguistics: Human Language Technologies, Volume 1 (Long
  Papers)}}. \bibinfo{publisher}{Association for Computational Linguistics},
  \bibinfo{address}{New Orleans, Louisiana}, \bibinfo{pages}{293--302}.
\newblock
\urldef\tempurl%
\url{https://doi.org/10.18653/v1/N18-1027}
\showDOI{\tempurl}


\bibitem[Sang and Meulder(2003)]%
        {entity3}
\bibfield{author}{\bibinfo{person}{Erik F. Tjong~Kim Sang} {and}
  \bibinfo{person}{Fien~De Meulder}.} \bibinfo{year}{2003}\natexlab{}.
\newblock \showarticletitle{Introduction to the CoNLL-2003 Shared Task:
  Language-Independent Named Entity Recognition}.
\newblock \bibinfo{journal}{\emph{CoRR}}  \bibinfo{volume}{cs.CL/0306050}
  (\bibinfo{year}{2003}).
\newblock
\urldef\tempurl%
\url{http://arxiv.org/abs/cs/0306050}
\showURL{%
\tempurl}


\bibitem[Sari et~al\mbox{.}(2010)]%
        {rule_entity}
\bibfield{author}{\bibinfo{person}{Yunita Sari}, \bibinfo{person}{Mohd~Fadzil
  Hassan}, {and} \bibinfo{person}{Norshuhani Zamin}.}
  \bibinfo{year}{2010}\natexlab{}.
\newblock \showarticletitle{Rule-based pattern extractor and named entity
  recognition: A hybrid approach}. In \bibinfo{booktitle}{\emph{2010
  International Symposium on Information Technology}},
  Vol.~\bibinfo{volume}{2}. \bibinfo{pages}{563--568}.
\newblock


\bibitem[Shukla et~al\mbox{.}(2022)]%
        {shukla-etal-2022-legal}
\bibfield{author}{\bibinfo{person}{Abhay Shukla}, \bibinfo{person}{Paheli
  Bhattacharya}, \bibinfo{person}{Soham Poddar}, \bibinfo{person}{Rajdeep
  Mukherjee}, \bibinfo{person}{Kripabandhu Ghosh}, \bibinfo{person}{Pawan
  Goyal}, {and} \bibinfo{person}{Saptarshi Ghosh}.}
  \bibinfo{year}{2022}\natexlab{}.
\newblock \showarticletitle{Legal Case Document Summarization: Extractive and
  Abstractive Methods and their Evaluation}. In
  \bibinfo{booktitle}{\emph{Proceedings of the 2nd Conference of the
  Asia-Pacific Chapter of the Association for Computational Linguistics and the
  12th International Joint Conference on Natural Language Processing (Volume 1:
  Long Papers)}}. \bibinfo{address}{Online only}.
\newblock


\bibitem[Skylaki et~al\mbox{.}(2020)]%
        {skylaki2020named}
\bibfield{author}{\bibinfo{person}{Stavroula Skylaki}, \bibinfo{person}{Ali
  Oskooei}, \bibinfo{person}{Omar Bari}, \bibinfo{person}{Nadja Herger}, {and}
  \bibinfo{person}{Zac Kriegman}.} \bibinfo{year}{2020}\natexlab{}.
\newblock \showarticletitle{Named entity recognition in the legal domain using
  a pointer generator network}.
\newblock \bibinfo{journal}{\emph{arXiv preprint arXiv:2012.09936}}
  (\bibinfo{year}{2020}).
\newblock


\bibitem[Sun et~al\mbox{.}(2018)]%
        {entity4}
\bibfield{author}{\bibinfo{person}{Peng Sun}, \bibinfo{person}{Xuezhen Yang},
  \bibinfo{person}{Xiaobing Zhao}, {and} \bibinfo{person}{Zhijuan Wang}.}
  \bibinfo{year}{2018}\natexlab{}.
\newblock \showarticletitle{An Overview of Named Entity Recognition}. In
  \bibinfo{booktitle}{\emph{2018 International Conference on Asian Language
  Processing (IALP)}}. \bibinfo{pages}{273--278}.
\newblock
\urldef\tempurl%
\url{https://doi.org/10.1109/IALP.2018.8629225}
\showDOI{\tempurl}


\bibitem[Tian et~al\mbox{.}(2020)]%
        {entity}
\bibfield{author}{\bibinfo{person}{Yuanhe Tian}, \bibinfo{person}{Wang Shen},
  \bibinfo{person}{Yan Song}, \bibinfo{person}{Fei Xia}, \bibinfo{person}{Min
  He}, {and} \bibinfo{person}{Kenli Li}.} \bibinfo{year}{2020}\natexlab{}.
\newblock \bibinfo{title}{Improving Biomedical Named Entity Recognition with
  Syntactic Information}.
\newblock
\newblock
\urldef\tempurl%
\url{https://doi.org/10.21203/rs.3.rs-21994/v1}
\showDOI{\tempurl}


\bibitem[Tran et~al\mbox{.}(2018)]%
        {catchphrase1}
\bibfield{author}{\bibinfo{person}{Vu~D. Tran}, \bibinfo{person}{Minh~Le
  Nguyen}, {and} \bibinfo{person}{Ken Satoh}.} \bibinfo{year}{2018}\natexlab{}.
\newblock \showarticletitle{Automatic Catchphrase Extraction from Legal Case
  Documents via Scoring using Deep Neural Networks}.
\newblock \bibinfo{journal}{\emph{CoRR}}  \bibinfo{volume}{abs/1809.05219}
  (\bibinfo{year}{2018}).
\newblock


\bibitem[Zhang et~al\mbox{.}(2014)]%
        {syntactic}
\bibfield{author}{\bibinfo{person}{Xiantao Zhang}, \bibinfo{person}{Dongchen
  Li}, {and} \bibinfo{person}{Xihong Wu}.} \bibinfo{year}{2014}\natexlab{}.
\newblock \showarticletitle{Parsing named entity as syntactic structure}. In
  \bibinfo{booktitle}{\emph{Interspeech}}.
\newblock
\urldef\tempurl%
\url{https://api.semanticscholar.org/CorpusID:33212550}
\showURL{%
\tempurl}


\end{thebibliography}
\end{document}